\documentclass[notoc]{JHEP3}

\pdfoutput=1

\usepackage{graphicx,multicol,multirow}
\usepackage{amsmath}

\newcommand\fverb{\setbox\fverbbox=\hbox\bgroup\verb}
\newcommand\fverbdo{\egroup\medskip\noindent%
			\fbox{\unhbox\fverbbox}\ }
\newcommand\fverbit{\egroup\item[\fbox{\unhbox\fverbbox}]}
\newbox\fverbbox

\renewcommand{\vec}[1]{\mathbf{#1}}

\title{Systematic effects in the extraction of the `WMAP haze'}
\author{Philipp Mertsch \& Subir Sarkar\\
Department of Physics, University of Oxford, 1 Keble Road, Oxford OX1 3NP, UK\\
E-mail: \email{p.mertsch@physics.ox.ac.uk}, \email{s.sarkar@physics.ox.ac.uk}}

\received{\today} 
\accepted{\today}

\abstract{ The extraction of a `haze' from the WMAP microwave skymaps
  is based on subtraction of known foregrounds, {\em viz.} free-free
  (bremsstrahlung), thermal dust and synchrotron, each traced by other
  skymaps. While the 408~MHz all-sky survey is used for the
  synchrotron template, the WMAP bands are at tens of GHz where the
  spatial distribution of the radiating cosmic ray electrons ought to
  be quite different because of the energy-dependence of their
  diffusion in the Galaxy. The systematic uncertainty this introduces
  in the residual skymap is comparable to the claimed haze and can,
  for certain source distributions, have a very similar spectrum and
  latitudinal profile and even a somewhat similar morphology. Hence
  caution must be exercised in interpreting the `haze' as a physical
  signature of, {\it e.g.,} dark matter annihilation in the Galactic
  centre.  }


\preprint{arXiv:1004.3056}

\begin{document}

\section{Introduction}

The measurement of anisotropies in the cosmic microwave background
(CMB) by COBE and WMAP has ushered in an exciting new era in
cosmology. The study of the cosmic signal requires careful subtraction
of galactic foreground emissions and this will become even more
crucial for studies of the `B-mode' polarisation signal by PLANCK and
the proposed CMBPol satellites \cite{Dunkley:2008am}. It is
interesting in this context that the subtraction of all known
foregrounds, {\it i.e.,} free-free (bremsstrahlung), thermal dust and
synchrotron (as well as the CMB), from the WMAP skymaps leaves an
anomalous emission --- the ``WMAP haze''
\cite{Finkbeiner:2003im}. This has a roughly spherical morphology
localised around the centre of the Galaxy, and a harder spectrum
\cite{Dobler:2007wv} than synchrotron radiation by relativistic cosmic
ray (CR) electrons from standard astrophysical sources like supernova
remnants (SNRs). An independent analysis has confirmed the existence
of the haze \cite{Bottino:2009uc}, but others do not find the evidence
to be significant \cite{Dickinson:2009yg,Cumberbatch:2009ji}.

It was believed initially that the haze is free-free emission from
ionised gas too hot to be traced by recombination line maps but too
cold to be visible in X-rays \cite{Finkbeiner:2003im}. However it was
suggested later that it is in fact synchrotron emission from a new
population of relativistic electrons,\footnote{Here and in the
  following we use ``electrons'' when referring to both electrons and
  positrons.} produced by dark matter annihilation
\cite{Finkbeiner:2004us}. It is indeed thus possible to explain the
haze \cite{Hooper:2007kb} although other authors argue that the
annihilation cross-section needs to be significantly boosted over the
usual estimate for thermal relic dark matter
\cite{Cumberbatch:2009ji}. There have also been attempts to fit both
the morphology and spectrum of the haze by ascribing it to electrons
emitted by pulsars with a hard spectrum \cite{Kaplinghat:2009ix,
  Harding:2009ye}; however the expected haze is then less spherical
since most pulsars are in the galactic disk. This is also true of SNRs
which have in fact recently been invoked \cite{Blasi:2009hv,
  Ahlers:2009ae} as sources of positrons with a hard spectrum to
explain the rise in the cosmic ray positron fraction at high energies
measured by PAMELA \cite{Adriani:2008zr}.

The presence of an additional population of relativistic electrons in
the galactic centre appears to be supported by a recent analysis
\cite{Dobler:2009xz} of the $\gamma$-ray sky as observed by Fermi-LAT
\cite{Atwood:2009ez}. It is argued that an excess over known
components is also present in $\gamma$-rays, most likely due to
inverse-Compton scattering (ICS) by relativistic electrons, and that
the underlying electron distribution is compatible with the WMAP haze
\cite{Dobler:2009xz}. While a signature in ICS is naturally expected
if there is indeed an additional population of electrons with a hard
spectrum, it was pointed out \cite{Linden:2010ea} that some template
maps applied in this analysis \cite{Dobler:2009xz} are in fact
inappropriate and underestimate both the $\pi^0$ decay and ICS
contributions to the $\gamma$-ray emission, in particular in the
galactic centre region. The analysis in Ref.~\cite{Dobler:2009xz}
using the Fermi diffuse model that is believed to be a better tracer
of $\pi^0$ decay and ICS, however, again shows a residual.  The Fermi
itself collaboration has not claimed any excess in the galactic centre
region over the standard diffuse $\gamma$-ray background
\cite{Abdo:2010nz,Casandjian:2009wq}.

A crucial ingredient of both studies
\cite{Finkbeiner:2003im,Bottino:2009uc} that identify a microwave haze
is the extrapolation of the morphology of the synchrotron radiation
template from 408 MHz to the WMAP bands at 23 (K), 33 (Ka), 41 (Q), 61
(V) and 94 (W) GHz, {\em i.e.} over two orders of magnitude in
frequency. In fact the spatial distribution of the radiating CR
electrons is likely to differ significantly given their energy
dependent diffusive transport in the Galaxy. Instead of attempting
such a bold extrapolation, other studies, including the analysis by
the WMAP collaboration \cite{Gold:2010fm}, employ the K-Ka difference
map as a tracer of synchrotron emission (despite some contamination by
free-free emission and an anomalous component which has been
interpreted (see, {\it e.g.,}\cite{deOliveiraCosta:2003az}) as
spinning dust \cite{Draine:1998gq}). However although both maps are
dominated by synchrotron radiation, such a template could also contain
any unidentified radiation, such as a possible haze, and therefore
cannot exclude it.

CR transport in the Galaxy is dominated by diffusion through
interstellar magnetic fields with an {\em energy-dependent} diffusion
coefficient $D(E) = D_0 E^{\delta}$ where $\delta = 0.3 \mathellipsis
0.7$ \cite{Strong:2007nh}. Taking the energy loss rate $b (E) =
\mathrm{d}E/ \mathrm{d}t = b_0 E^2$ as is appropriate for synchrotron
and ICS, the diffusion length $\ell$ is
%
\begin{equation*}
\ell(E) \approx 5 \left(\frac{E}{\text{GeV}}\right)^{(\delta-1)/2}\,\text{kpc}\,,
\end{equation*}
for the standard values $D_0 = 10^{28}\,\text{cm}^{2}\text{s}^{-1}$
and $b_0 = 10^{-16}\,\text{s}^{-1}$ \cite{Strong:2007nh}. Therefore,
the distance that GeV energy electrons can diffuse is comparable to
the Kpc scale on which the source distribution varies; moreover it
changes by a factor of 2.4 (1.5) for $\delta = 0.3$ (0.7) in the
energy range $\sim4-50$ GeV (corresponding to peak synchrotron
frequencies between 408 MHz and 50 GHz for a magnetic field of
$6\,\mu\text{G}$).  As a consequence the $\sim 50 \, \text{GeV}$
electrons will trace the source distribution much better than the
$\sim 4 \, \text{GeV}$ electrons which diffuse further away from the
sources and wash out their distribution.  The synchrotron map at 408
MHz {\em cannot} therefore be a good tracer of synchrotron radiation
at much higher, in particular WMAP, frequencies. Relying on such a
crude extrapolation of the morphology of synchrotron emission can thus
potentially introduce unphysical residuals. We estimate these by
simulating synchrotron skymaps at 408 MHz and the WMAP frequencies and
feeding these into the template subtraction process
\cite{Dobler:2007wv}. We show that this leads to residuals of the same
order as the claimed haze, which can in fact be matched in spectrum
and latitudinal profile for a particular source distribution in the
galactic disk. We conclude therefore that the WMAP haze might be an
artifact of inappropriate template subtraction rather than evidence of
an exotic origin, e.g. dark matter annihilation.

\section{Template subtraction}
\label{sec:TemplateSubtraction}

The subtraction method is based on a multilinear regression of the CMB
subtracted WMAP data using foreground templates for free-free (f),
dust correlated (d) and synchrotron emission (s). Technically this can
be achieved by assembling the maps represented each by a vector of all
pixels, that is $\vec{f}$, $\vec{d}$ and $\vec{s}$, into one `template
matrix': $P = \left(\vec{f},\vec{d}, \vec{s}\right)$. (The template
for the haze, $\vec{h}$, is appended later.) The pseudo-inverse of
$P$, $P^+$, allows to determine the coefficients $\vec{a} = P^+
\vec{w}$ that minimise the $\chi^2 = || \vec{w} - P\,\vec{a} ||^2
/\sigma^2$ for the different templates at the WMAP frequencies;
$\sigma$ is the mean measurement noise in each frequency band. For
details see Ref. \cite{Dobler:2007wv}.

Since we are interested only in the effect of the electron diffusion
on the subtraction of the synchrotron foreground we do not use the
free-free and dust templates or radio skymaps that are strongly
affected by local structures such as Loop I
\cite{Berkhuijsen:1971}. Instead we simulate both the synchrotron
skymap at 408 MHz and the skymaps in the WMAP frequency range with the
{\tt GALPROP} code \cite{Moskalenko:1997gh}. We adopt the same mask as
in Ref. \cite{Dobler:2007wv} which excises pixels along the galactic
plane, around radio sources and in directions of excessive absorption.

To allow comparison with the results of Ref. \cite{Dobler:2007wv} we
apply the same fitting procedure over the whole sky. In order to
determine the magnitude of the `haze' we append a template
$\vec{h}=(1/\theta -1/\theta_0)$ to $P$ where $\theta = \sqrt{\ell^2 +
  b^2}$ is in galactic coordinates and $\theta_0 = 45^\circ$. This
corresponds to the ``FS8'' fit performed in \cite{Dobler:2007wv} and
adding the haze back to the residual maps gives the ``FS8 + haze''
maps. We determine the latitudinal profile of the residual for $\ell =
0^\circ$ south of the galactic centre direction. As our simulated maps
do not contain any localised structures, we do not need to divide the
sky into several regions and fit them independently, as was done with
the ``RG8'' fit \cite{Dobler:2007wv} . We have checked explicitly that
doing so does not change the profiles of the residual intensity or the
spectral indices.

We have checked that our procedure gives a residual `haze' in
agreement with Ref. \cite{Dobler:2007wv} when we subtract the 408 MHz
survey, the H$\alpha$ and dust skymaps from the WMAP skymaps. Although
with the CMB estimator ``CMB5'' we find a residual intensity of the
same magnitude at 23 GHz, its spectral index of about $-0.7$ is
somewhat softer than in Ref. \cite{Dobler:2007wv}.

\section{Diffusion model}
\label{sec:diffusion_model}

The transport of CR electrons is governed by a diffusion-convection
equation \cite{Ginzburg:1990sk},
\begin{align*}
\frac{\partial n}{\partial t} =& \vec{\nabla}  \cdot 
\left( D_{xx} \vec{\nabla} n - \vec{v} \,n \right) + 
\frac{\partial}{\partial p} p^2 D_{pp} 
\frac{\partial}{\partial p} \frac{1}{p^2} n \\
&- \frac{\partial}{\partial p} \left( \dot{p} \, 
n - \frac{p}{3} \left( \vec{\nabla} \cdot \vec{v} \right) n \right) + q \,,
\end{align*}
where $n \, \mathrm{d}p$ is the number density of electrons with
momentum in $\left[p, p + \mathrm{d} p\right]$, $D_{xx} = D_{0 xx}
(p/4 \, \text{GeV})^\delta$ is the spatial diffusion coefficient,
$\vec{v}$ is the convection velocity, $D_{pp}$ is the momentum
diffusion coefficient and $q$ is the source power density. This
equation is numerically solved with the {\tt GALPROP} code {\tt
  v50.1p} in two dimensions, that is assuming azimuthal symmetry
around the galactic centre and enforcing the boundary condition $n
\equiv 0$ on a cylinder of radius $R = 20\,\text{kpc}$ and half-height
$z_\text{max}$ (see below).

The source power density $q$ factorises into a source energy spectrum
$q_0 E^{-\alpha}$ and a spatial variation $\sigma(r)
\mathrm{e}^{-z/z_\text{scale}}$ with $z_\text{scale} =
0.2\,\text{kpc}$. For the radial part we consider two
possibilities. The distribution of SNRs is expected to be correlated
with that of pulsars which is inferred by Lorimer to be
\cite{Lorimer:2003qc}
\begin{equation}
\label{eqn:Lorimer}
\sigma_\text{Lorimer} (r) = 64.6 \left(\frac{r}{\text{kpc}}\right)^{2.35} 
\mathrm{e}^{-r/1.528 \,\text{kpc}}\,.
\end{equation}
However, the determination of pulsar distances from their rotation
measures relies on knowledge of the thermal electron density
throughout the Galaxy and different distributions lead to different
functional forms for the inferred radial variation of the pulsar
density \cite{Lorimer:2006qs}. Therefore we also consider an
exponential source distribution
\begin{equation}
\label{eqn:exponential}
\sigma_\text{exp}(r) = \sigma_0 \mathrm{e}^{-r/2\,\text{kpc}}\,,
\end{equation}
following Refs.\cite{Paczynski:1990} and \cite{Sturner:1996}.

The normalisation $D_{0 xx}$, the scale height $z_\text{max}$ of the
CR halo and the spectral index $\delta$ of the diffusion coefficient
are usually determined from measurements of CR nuclei and nuclear
secondary to primary ratios. The measurement of CR `chronometers' like
${}^{10}\text{Be}/{}^9\text{Be}$ is still not precise enough to break
the degeneracy between $D_{0 xx}$ and $z_\text{max}$, so we vary
$z_\text{max}$ between $4\,\text{kpc}$ and $8\,\text{kpc}$ and vary
$D_{0 xx}$ only a little, checking that we have rough agreement with
the measured fluxes of nuclei and nuclear secondary-to-primary
ratios. On theoretical grounds \cite{Ptuskin:2005ax} one expects a
spectral break in the diffusion coefficient at $\approx 1 \,
\text{GeV}$. We fix the break energy to 1 GeV and vary $\delta_1$ and
$\delta_2$ below and above the break (keeping $\delta_1\geq
\delta_2$), again trying to satisfy all local CR measurements.

The source electron spectrum is usually assumed to have a break around
4~GeV \cite{Strong:2004de} so we fix the electron source normalisation
$q_0 \sigma_0$ and the spectral indices $\alpha_1$ and $\alpha_2$
below and above the break by fitting the propagated flux to the
electron spectrum as measured at Earth
\cite{Strong:2009xp,Abdo:2009zk}. We apply Solar modulation in the
spherical approximation \cite{Gleeson:1968} with a median potential of
$\phi = 550\,\text{MV}$. Reacceleration and convection play a role at
energies below $10\,\text{GeV}$ and are therefore important for the
408 MHz map. For the Alfv\`en velocity $v_\mathrm{A}$ which determines
the strength of reacceleration via $D_{pp} \propto v_\mathrm{A}^2$ we
consider the range $0-50\,\text{km}\,\text{s}^{-1}$. {\tt GALPROP}
assumes the convection velocity to vary linearly with distance from
the galactic plane and we vary the slope ${\rm d} v_{\text{conv}}/{\rm
  d} z$ between 0 and $20\,\text{km}\,\text{s}^{-1}\,\text{kpc}^{-1}$.

\TABLE[b]{
\begin{tabular*}{0.62\columnwidth}{ c |  c | c }
\hline\hline 
&	Lorimer  &	exponential \\
\hline
Source & \multirow{2}{*}{{\em c.f.} eq. \ref{eqn:Lorimer}} 
& \multirow{2}{*}{{\em c.f.} eq. \ref{eqn:exponential}} \\ [-1ex]
distribution & & \\
$\alpha_1, \alpha_2$ 	& 1.2, 2.2	& 1.2, 2.2	\\
$D_{0 xx}$ 
& $5.75 \times 10^{28}\,\text{cm}^2\,\text{s}^{-1}$  
& $5.75 \times 10^{28}\,\text{cm}^2\,\text{s}^{-1}$ \\
$z_\text{max}$ 	& 4\,\text{kpc}	& 8\,\text{kpc}	\\
$\delta_1, \delta_2$ 	& 0.34, 0.34	& 0.1, 0.4	\\
$v_{\text{A}}$			& $50  \, \text{km}\,\text{s}^{-1}$ 
& $36  \, \text{km}\,\text{s}^{-1}$  \\
${\rm d} v_{\text{conv}}/{\rm d} z$ 
& $10 \, \text{km}\,\text{s}^{-1} \, \text{kpc}^{-1}$	
& $15 \, \text{km}\,\text{s}^{-1} \, \text{kpc}^{-1}$ \\
$B_0$ 	& $6.3 \, \mu \text{G}$	& $6.8 \, \mu \text{G}$ \\
$\rho$ 	& $5 \, \text{kpc}$ 	& $50 \, \text{kpc}$ \\
\hline\hline
\end{tabular*}
\caption{Parameters of source and diffusion models.}
\label{tbl:parameters}
}

Since the random component of the galactic magnetic field is known to
dominate over the regular component \cite{Beck:2008eb}, we neglect the
latter. For the radial dependence we adopt the usual exponential
fall-off where the radial scale $\rho$ and the (perpendicular
component of the) field strength $B_0$ at the galactic centre are
chosen to reproduce the 408 MHz skymap \cite{Haslam:1982}.  Although
it was initially believed \cite{Strong:1998fr} that an exponential
dependence on $z$ could give a satisfactory fit to the 408 MHz
latitude profile, the galactic field model was later refined
\cite{Orlando:2009xg} by considering different, non-exponential
behaviours which in fact give better fits.  We therefore apply the
method described in Ref. \cite{Philipps:1981} of determining the
emissivity dependence on $r$ for galactic longitude $\ell = \pm
180^\circ$ (towards the galactic anti-centre). With an estimate for
the electron density this translates into a $z$-dependence of the form
$a + b \exp{[(-|z|/\xi)^\kappa]}$ and this is iterated to convergence
where we find $a/b = 0.27$, $\xi = 0.51$ and $\kappa = 0.68$.

\section{Results}
\label{sec:Results}

\subsection{Lorimer source distribution}

The parameters of the diffusion model, the magnetic field and the
electron source spectrum have been adjusted as described above and the
values are shown for the Lorimer source distribution
(\ref{eqn:Lorimer}) in Table~\ref{tbl:parameters}. Almost all
diffusion model parameters are in the range also found by previous
{\tt GALPROP} studies
\cite{Strong:1998fr,Strong:2004de,Ptuskin:2005ax} to give a consistent
picture of GCRs. The only difference is the electron injection
spectrum which is chosen considerably harder to reproduce the new
measurement by Fermi-LAT at $\mathcal{O}(100) \, \text{GeV}$ energies
which was not available for the above mentioned studies.

\FIGURE[t]{
\includegraphics[scale=0.5]{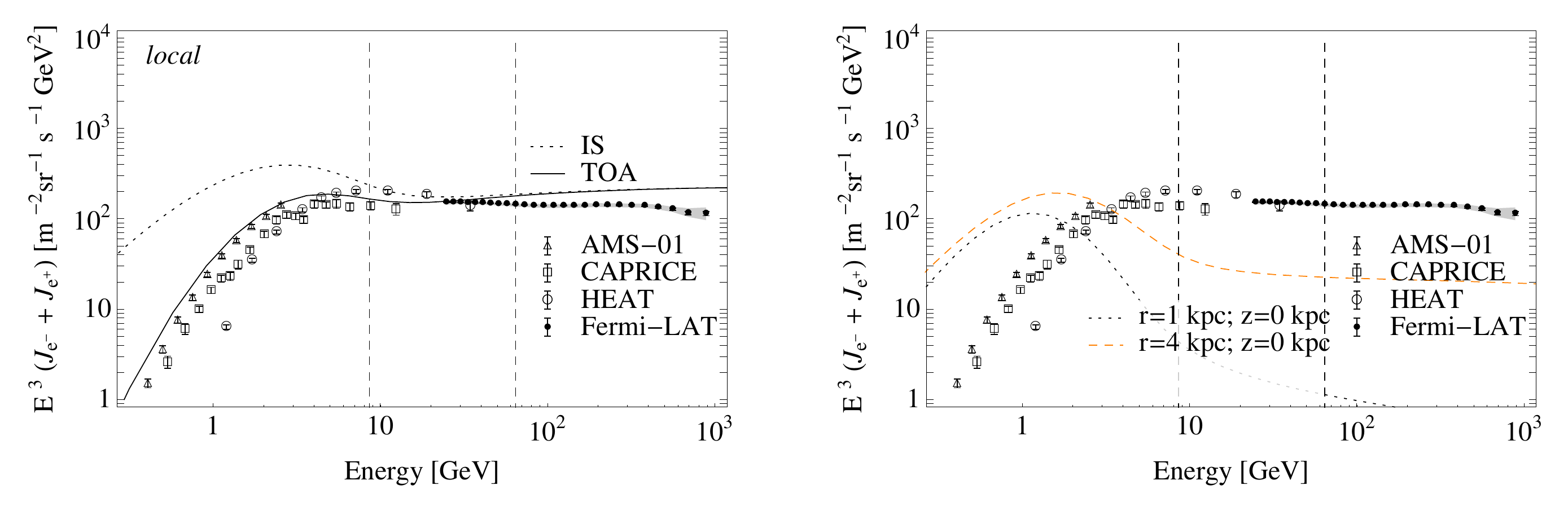}
\caption{{\bf Left panel.} The electron (plus positron) flux measured
  locally by AMS-01, CAPRICE, HEAT \cite{Strong:2009xp} and Fermi-LAT
  \cite{Abdo:2009zk}, compared with the expectation for the Lorimer
  source distribution (\ref{eqn:Lorimer}); the dotted line is the
  calculated interstellar (IS) flux while the solid line is its Solar
  modulated (TOA) value (with $\phi = 550$~MV). The dashed vertical
  lines show the energy corresponding to peak synchrotron frequencies
  of 408 MHz and 23 GHz for the local magnetic field.  {\bf Right
    panel.} The calculated electron (plus positron) flux at the
  positions $\{ (r, z) \} = \{ (1, 0), (4, 0)) \}$ (in kpc).}
\label{fig:pElPlusPos_50p_901019}
}

\FIGURE[!bh]{
\includegraphics[width=\textwidth]{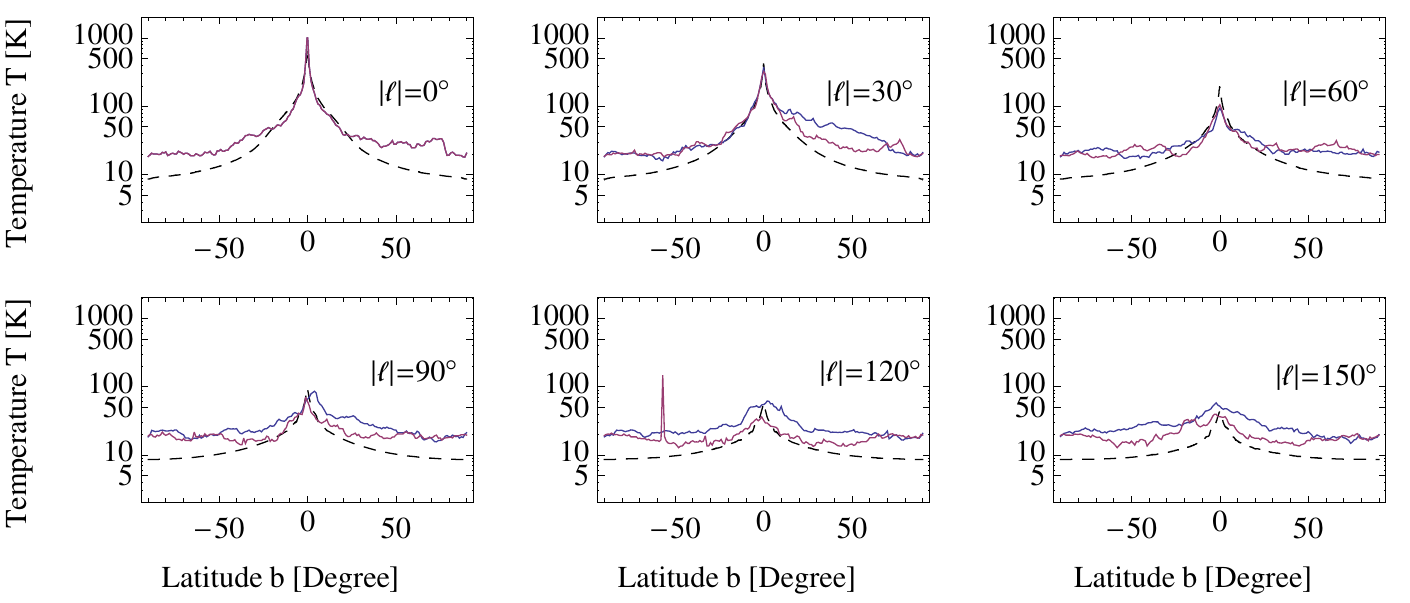}
\caption{The calculated latitudinal profile of galactic synchrotron
  radiation at 408 MHz (black dashed line) for galactic longitudes
  $|\ell|~=~0^\circ, 30^\circ, 60^\circ, 90^\circ, 120^\circ$ and
  $150^\circ$. The red (blue) solid line is the observed profile
  \cite{Haslam:1982} for positive (negative) $\ell$.}
\label{fig:pLats_50p_901019}
}

\FIGURE[t]{
\includegraphics[scale=1]{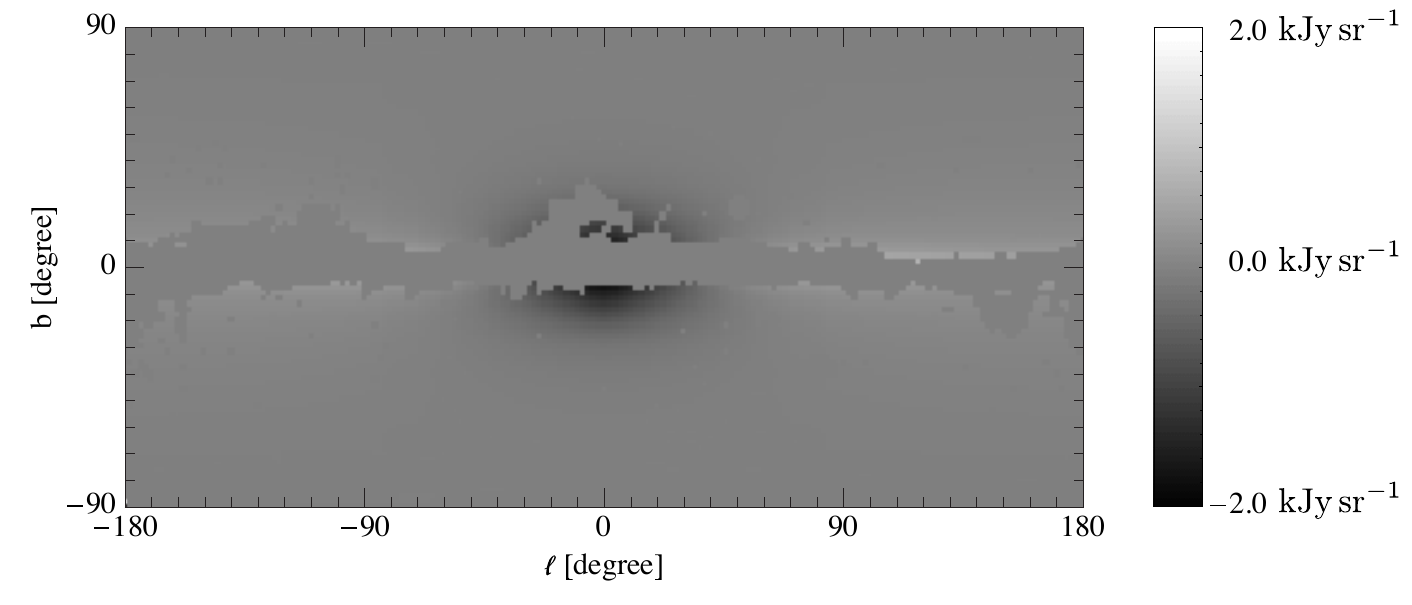}
\caption{Residual skymap in galactic coordinates for the Lorimer
  source distribution (\ref{eqn:Lorimer}).}
\label{fig:residual_skymap_Lorimer}
}

\FIGURE[h]{
\includegraphics[width=0.45\textwidth]{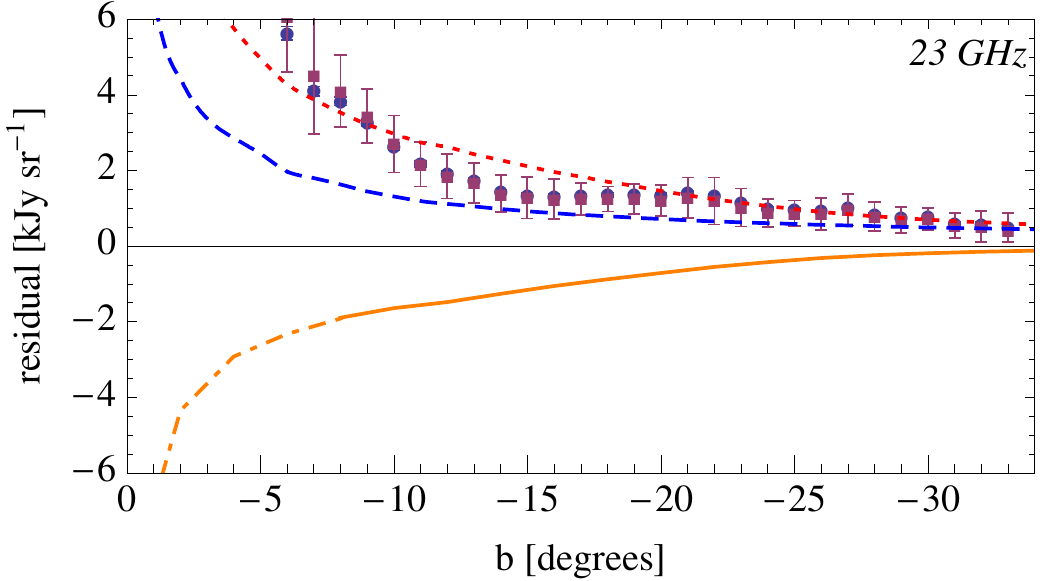}
\includegraphics[width=0.45\textwidth]{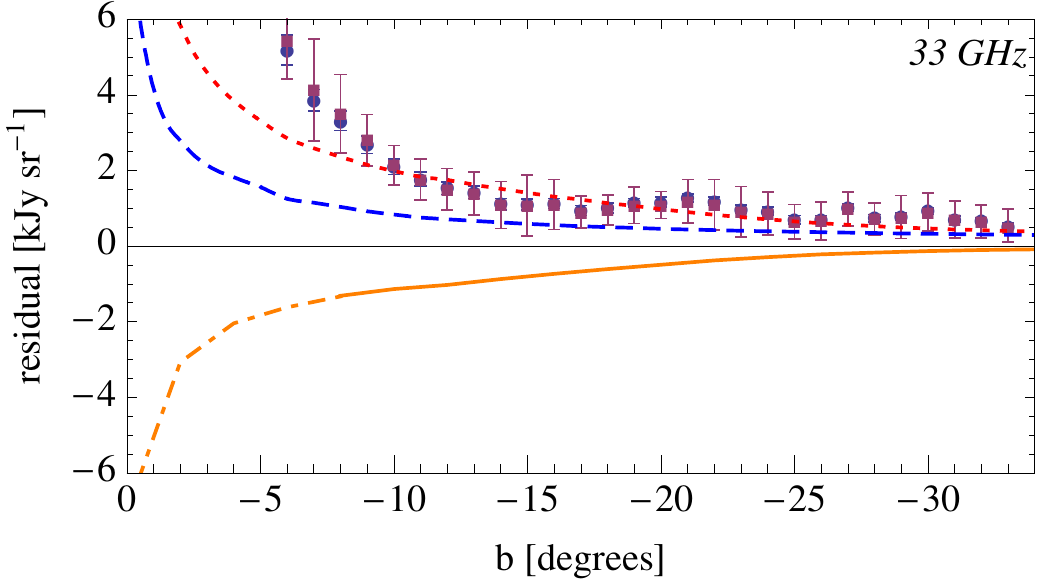}
\caption{Latitudinal profile of the K band residual outside (solid
  curve) and under (dot-dashed curve) the mask at 23 GHz (left panel)
  and 33 GHz (right panel) for $\ell = 0$. The square (circle) data
  points are the `haze' as extracted in Ref. \cite{Dobler:2007wv}
  (\cite{Hooper:2007kb}). The dotted line shows the extrapolated
  emission at 23 (33) GHz from scaling the simulated 408 MHz emission
  and the dashed line shows the actual simulated 23 (33) GHz
  emission.}
\label{fig:residual_profile_Lorimer}
}

The electron flux measured locally and at the positions $\{ (r, z) \}
= \{ (1, 0), (4, 0) \}$ (in kpc) are shown in
Fig. \ref{fig:pElPlusPos_50p_901019}. We note that close to the
galactic centre the electron flux responsible for synchrotron
radiation at 408 MHz is not only much softer but also suppressed by
over an order of magnitude with respect to its locally measured value.
The predicted local electron flux seems to be too hard at energies
above tens of GeV and overshoots the measurement by Fermi-LAT. We note
however that this is only an effect of the assumed \emph{continuous}
source distribution as implemented in the {\tt GALPROP} code by
default. At energies $\mathcal{O}(100) \, \text{GeV}$, however, where
the diffusion-loss length of electrons $\ell(E)$ becomes smaller than
the distance between the solar system and the nearest source(s), the
discreteness of GCR sources starts to play a role, leading to a
cut-off in the power law local electron spectrum (see
\cite{Ahlers:2009ae} for an illustration of this effect). We note that
the locally measured electron spectrum (from a discrete distribution
of sources) does in fact correspond to a power law spectrum $\propto
E^{-3}$ or slightly harder. This is in agreement with the spectrum
predicted from both our source distributions and injection spectra.

Fig. \ref{fig:pLats_50p_901019} shows the latitudinal profiles of the
synchrotron radiation at 408 MHz; in general, the fit is good for $b
\lesssim 50^\circ$ but underestimates the emission at larger
latitudes. It has been shown \cite{Philipps:1981} that this can
potentially be overcome by increasing the scale height of the
synchrotron emissivity at larger galactic radii. The remaining
discrepancies between the simulated and measured profiles are probably
due to the assumption of rotational symmetry. This leads to an
underestimation of the synchrotron radiation along tangents of the
spiral arms and an overestimation between them. For example, the
Carina arm is tangent at $75^\circ$ and the Sagittarius arm at
$-40^\circ$, so both $\ell = +60^\circ$ and $\ell = -60^\circ$ are
between spiral arms and thus slightly overestimated, in particular in
the galactic plane. It is also clear that point sources (that have not
been subtracted from the 408 MHz data) are not accounted for in our
calculation ({\em e.g.,} Fornax at $\ell \simeq 120^\circ$, $b \simeq
-57^\circ$).

The skymap of the residual $r(\ell,b)$
(Fig. \ref{fig:residual_skymap_Lorimer}) shows a deficit for $|\ell|
\leq 40^\circ$ and $|b| \leq 20^\circ$. Further away from the galactic
centre direction there is a slight excess. The residual specific
intensity (Fig. \ref{fig:residual_profile_Lorimer}) is of opposite
sign but its absolute value is of the same order of magnitude as the
`haze' at 23 and 33 GHz.  Such \emph{intrinsic} residuals of the
template subtraction can substantially modify the magnitude,
morphology and/or spectrum of any \emph{physical} residual that might
be present in the microwave data. We emphasise that this will have
important consequences for the allowed parameter space of models
trying to explain such a potential excess, for example by dark matter
annihilation.

\subsection{Exponential source distribution}

For the exponential source distribution (\ref{eqn:exponential}), the
electron fluxes are shown in
Fig. \ref{fig:pElPlusPos_50p_910025b}. Close to the galactic centre,
it is larger by about an order of magnitude than measured locally and
slightly harder. The latitudinal profiles of the synchrotron radiation
at 408 MHz are shown in Fig.  \ref{fig:pLats_50p_910025b}.

\FIGURE[h]{
\includegraphics[width=\textwidth]{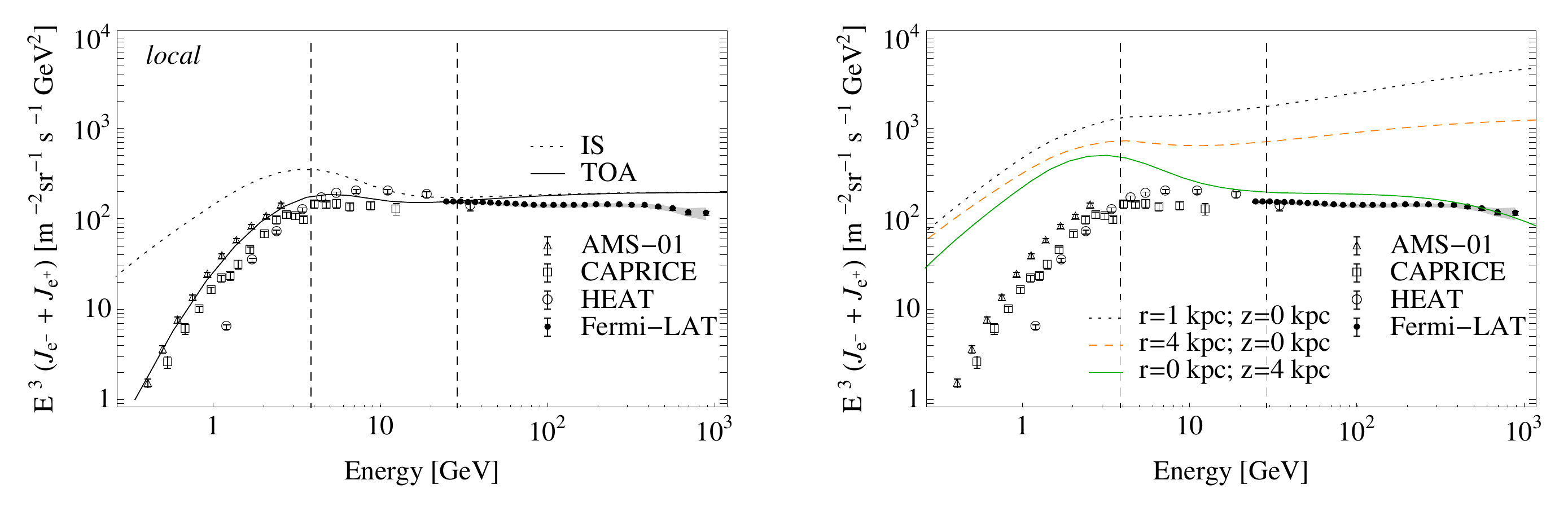}
\caption{Same as in Fig. \ref{fig:pElPlusPos_50p_901019}, but for the
  exponential source distribution (\ref{eqn:exponential}).}
\label{fig:pElPlusPos_50p_910025b}
}

\FIGURE[b]{
\includegraphics[width=\textwidth]{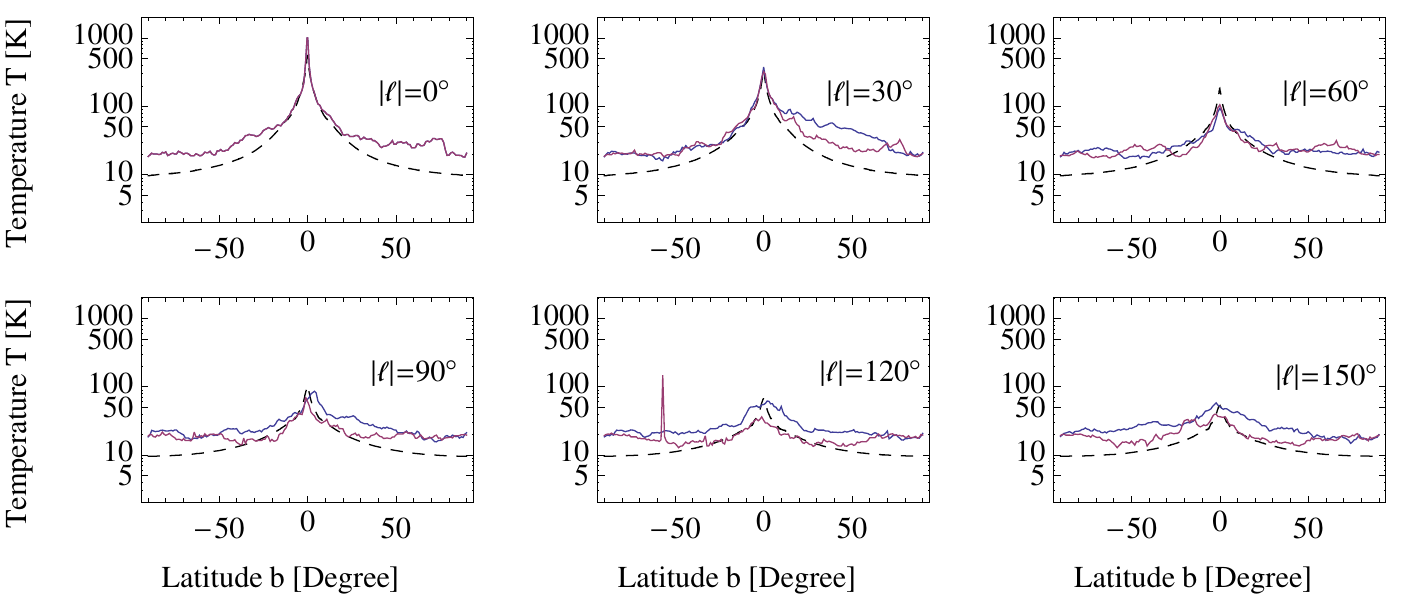}
\caption{Same as in Fig. \ref{fig:pLats_50p_901019}, but for the
  exponential source distribution (\ref{eqn:exponential}).}
\label{fig:pLats_50p_910025b}
}

The residual skymap contains a roughly spherical excess around the
centre of the map, although somewhat more extended in longitude than
in latitude (see Fig. \ref{fig:residual_skymap_exponential}).

We note that the systematic uncertainty of the residual intensity (as
determined from real skymaps) induced by chance correlations between
the `haze' template and the CMB has been estimated in
Ref. \cite{Dobler:2007wv} and can be read off their Fig. 8 as $\pm
11.8 \, \vec{h} \, \text{kJy} \, \text{sr}^{-1}$ ($\pm 23.7 \, \vec{h}
\, \text{kJy} \, \text{sr}^{-1}$) in the 23 GHz (33 GHz) band. We
therefore allow for an offset of our calculated residual relative to
the `haze' template in this range when fitting the residuals from real
skymaps.  The residual intensity
(Fig. \ref{fig:residual_profile_exponential}) {\em matches} the
claimed WMAP haze in latitudinal profile.

\FIGURE[t]{
\includegraphics[scale=1]{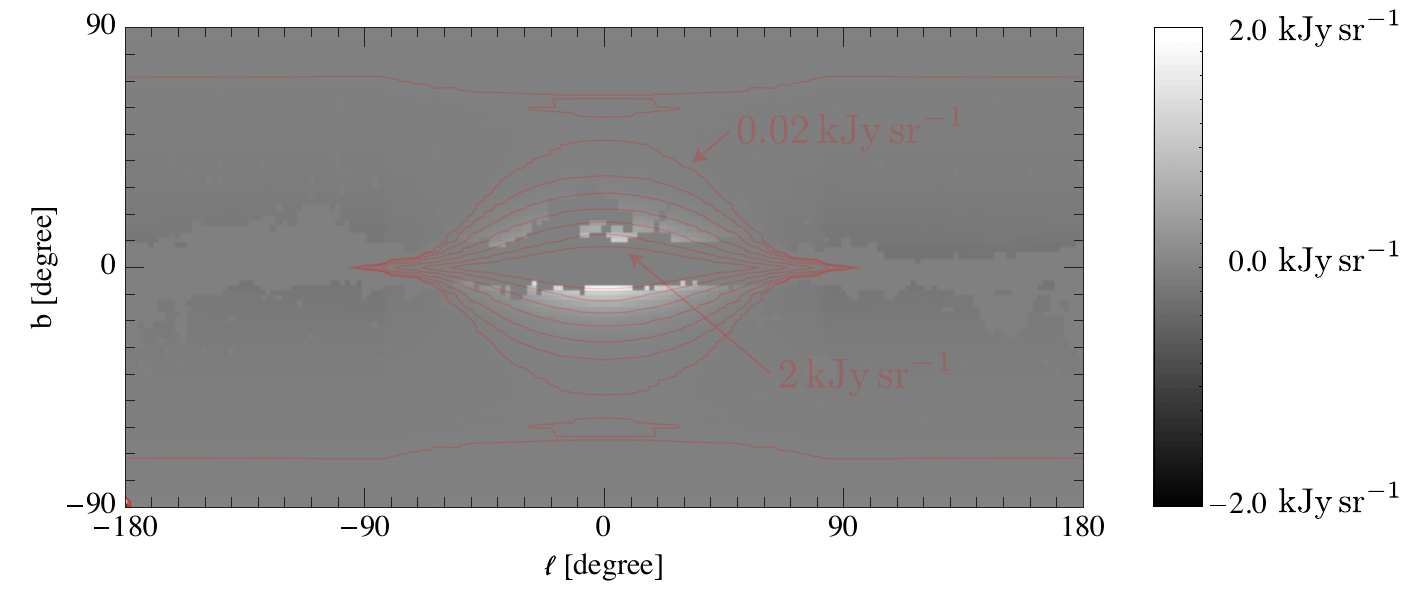}
\caption{Residual skymap in galactic coordinates for the exponential
  source distribution (\ref{eqn:exponential}). The contour lines are
  equally spaced between $0.02 \, \text{kJy} \, \text{sr}^{-1}$ and $2
  \, \text{kJy} \, \text{sr}^{-1}$.}
\label{fig:residual_skymap_exponential}
}

\FIGURE[t]{
\includegraphics[width=0.45\textwidth]{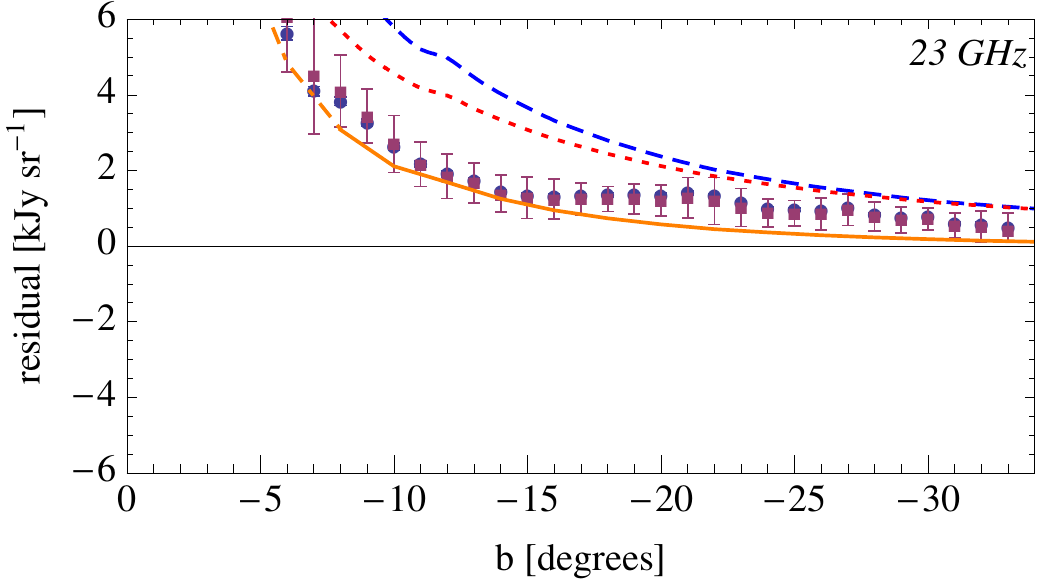}
\includegraphics[width=0.45\textwidth]{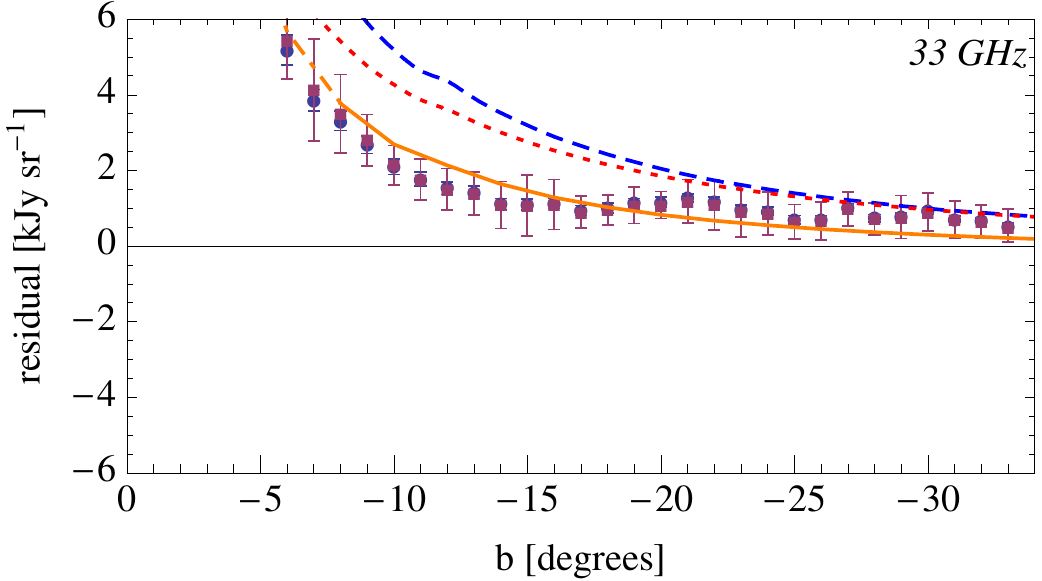}
\caption{Same as in Fig. \ref{fig:residual_profile_Lorimer}, but for
  the exponential source distribution (\ref{eqn:exponential}). We have
  added an offset to the calculated residual of $+11.8 \, \vec{h} \,
  \text{kJy} \, \text{sr}^{-1}$ ($+23.7 \, \vec{h} \, \text{kJy} \,
  \text{sr}^{-1}$) in the 23 GHz (33 GHz) band reflecting the
  systematic uncertainty from chance correlations between the `haze'
  template and the CMB.}
\label{fig:residual_profile_exponential}
}

\FIGURE[h!]{
\vspace{0cm} \includegraphics[scale=1]{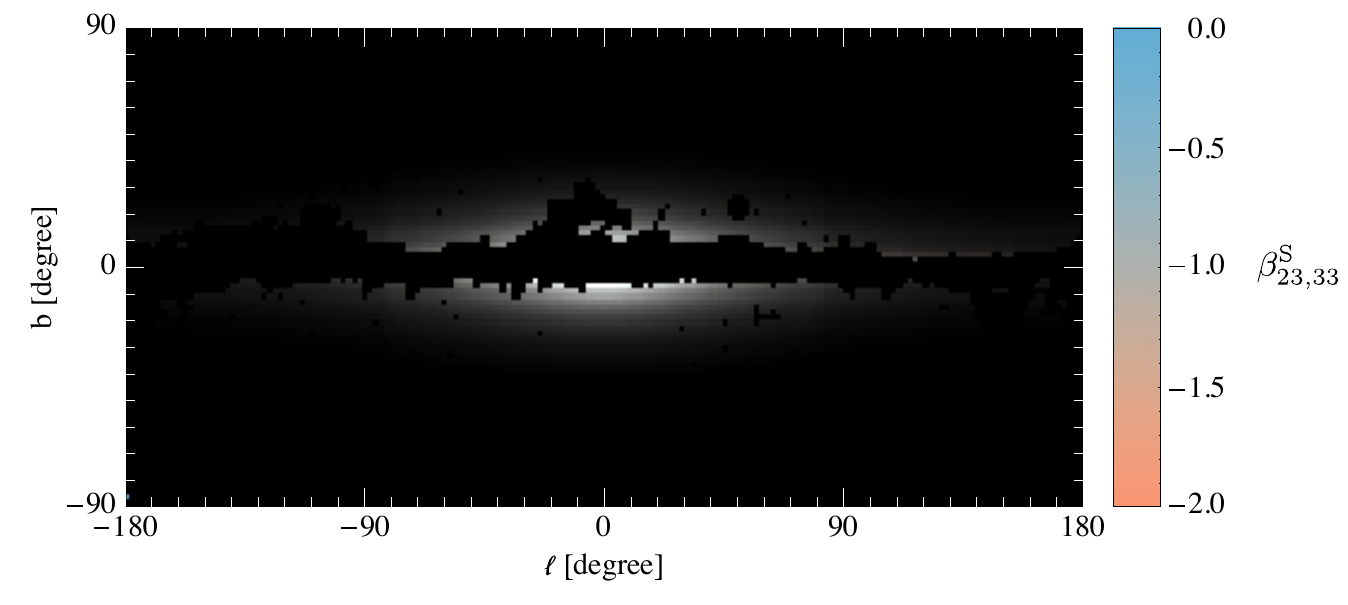} \\
\vspace{0cm} \includegraphics[scale=1]{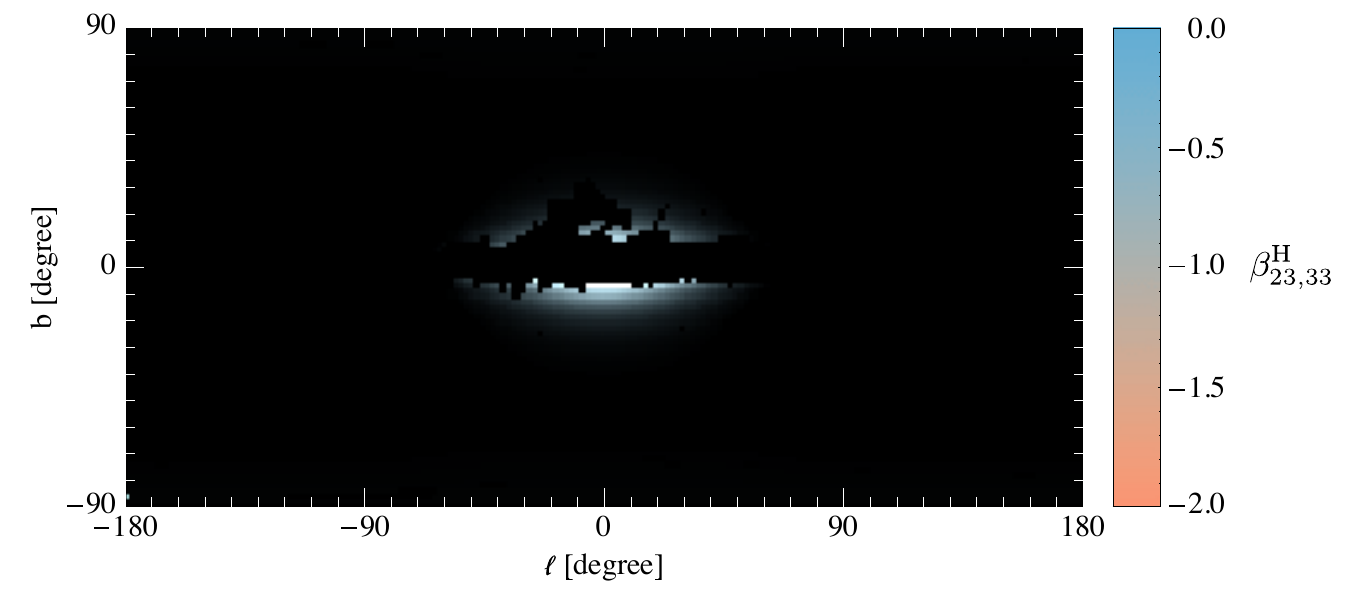} \\[-0.6cm]
\caption[]{Colour maps of spectral indices between 23 and 33 GHz
  defined in eq. \ref{eqn:DefSpecInd} scaled by the 23 GHz intensity
  for synchrotron + residual (top panel) and residual only (bottom
  panel).}
\label{fig:SpecIndMaps}
}

\FIGURE[h!]{
\includegraphics[scale=0.8]{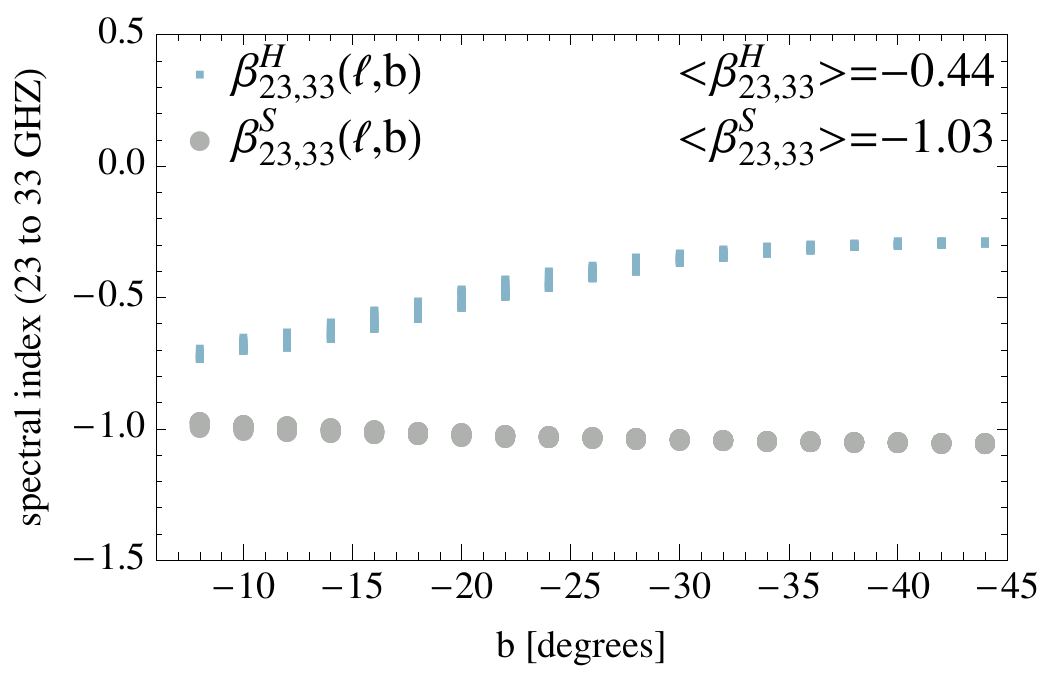}
\caption[]{Spectral indices of the unmasked pixels in the region south
  of the galactic centre ($b \in [-45 {}^\circ, 0{}^\circ], \, \ell
  \in [-25{}^\circ, 25{}^\circ]$) as a function of latitude for
  residual + synchrotron (large beige circles) and residual alone
  (small blue squares). The average spectral indices, $\beta_\text{S}$
  and $\beta_\text{H}$, are shown in the upper right corner.}
\label{fig:SpecIndLat}
}

To compare our results to those of Ref. \cite{Dobler:2007wv}, we also
determine the average spectral index (for details see Appendix
\ref{sec:app}) in a region south of the galactic centre, $b \in [-45
  {}^\circ, 0{}^\circ], \, \ell \in [-25{}^\circ, 25{}^\circ]$. The
colour maps of spectral indices scaled by intensity are shown in
Fig. \ref{fig:SpecIndMaps}, both for the synchrotron + residual and
for the residual alone. Not only is the synchrotron emission much more
disk-like than the residual, but the spectral index of the residual is
also considerably harder than the synchrotron spectral index. This is
to be compared with Fig. 7 of Ref. \cite{Dobler:2007wv} which exhibits
the same qualitative behaviour.

Furthermore, we show the spectral index for the unmasked pixels in the
region south of the galactic centre (as defined above) as a function
of latitude in Fig. \ref{fig:SpecIndLat}, again both for the
synchrotron + residual and for the residual alone. With average
indices of $\langle \beta_{23,33}^\text{H} \rangle = -0.44$ for the
residual and of $\langle \beta_{23,33}^\text{S} \rangle = -1.03$ for
the residual + synchrotron in this region, we find that the residual
index is harder than the synchrotron index by 0.6, which is in
excellent agreement with the findings of
Ref. \cite{Dobler:2007wv}. The values of our different model
parameters are shown in Table \ref{tbl:parameters}.

\section{Discussion}

\FIGURE[t]{
\includegraphics[scale=1]{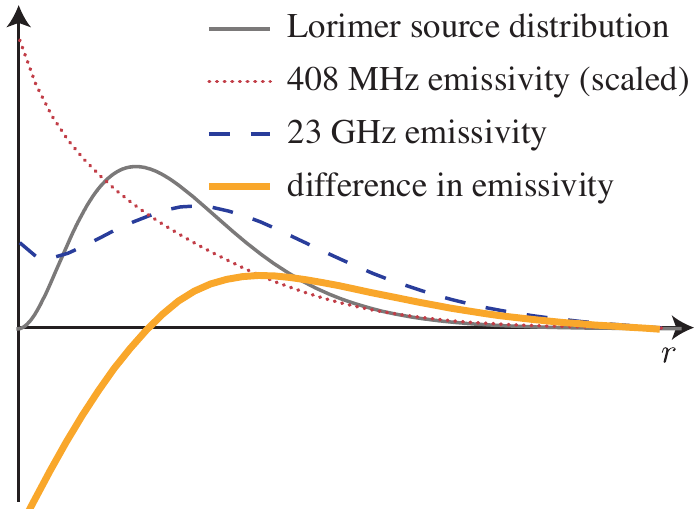}
\includegraphics[scale=1]{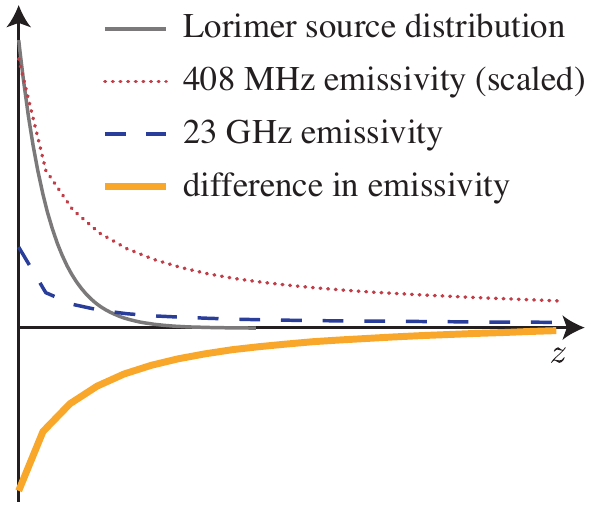} \\
\caption[]{The (scaled) synchrotron emissivity at 408~MHz and 23~GHz,
  and their difference for the Lorimer type source distribution at $z
  = 0 \, \text{kpc}$ (left) and $r = 0 \, \text{kpc}$ (right).}
\label{fig:differenceLorimer}
}

\FIGURE[h]{
\includegraphics[scale=1]{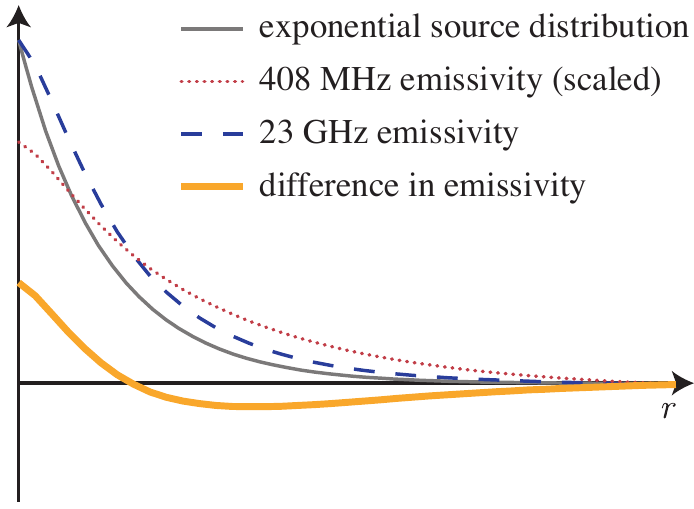}
\includegraphics[scale=1]{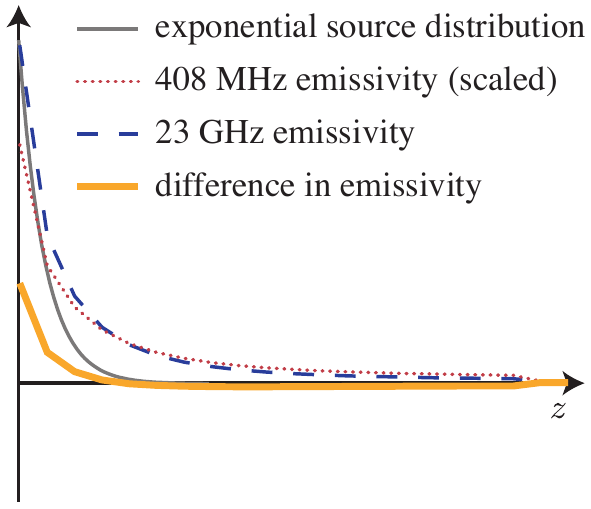}
\caption[]{Same as Fig. \ref{fig:differenceLorimer} but for the
  exponential source distribution.}
\label{fig:differenceExponential}
}

To qualitatively understand these results, we consider the
longitudinal and latitudinal profiles of the synchrotron intensity
$I(\ell, b)$; for simplicity let us constrain ourselves to the
galactic plane, {\it i.e.,} $z \equiv 0$ and a plane perpendicular,
{\it i.e.,} $r \equiv 0$. The intensity in any direction $\ell, b$ is
given by the integral of the synchrotron emissivity over the line of
sight and this samples the radial distribution of the relativistic
electron density in the range $r \in [d \sin\ell, \,R]$ and $z \in [0,
  \min[z_\text{max},(d + R) \tan b] ]$, where $d$ is the distance of
the Sun from the galactic centre. Since the fitting procedure
minimises the square of the difference in the maps, the sign and size
of the residual is determined not by the absolute difference but by
the difference in the {\em slopes} in $r$ and $z$ of the emissivity
$\varepsilon(r,z)$ at 408~MHz and the WMAP frequencies. The difference
in the slopes reflects the energy dependence of the electron diffusion
--- higher energy electrons lose their energy more quickly, hence
their emissivity traces the source distribution more closely than does
the emissivity of low energy electrons.

Considering the galactic plane ($z \equiv 0$) first, for the pulsar
source distribution the low energy electrons peak at the galactic
centre whereas the high-energy electrons peak further away along the
galactic plane (see Fig. \ref{fig:differenceLorimer}). This leads to a
deficit for small radii (translating to small longitudes) and a slight
excess further away from the galactic centre (see also
Fig. \ref{fig:residual_skymap_Lorimer}). For the exponential source
distribution the radial distribution of synchrotron emissivity is
steeper at higher energies. The template subtraction therefore yields
a residual with an excess around the centre direction and a deficit
further away along the galactic plane (see
Fig. \ref{fig:differenceExponential}).

Perpendicular to the galactic plane ($r \equiv 0$), the slope of the
emissivities of low and high energy electrons is similar for both
models, since the $z$-dependence of the source distributions is the
same. However, due to the different variation with galacto-centric
radius, the relative normalisations $\vec{a}$ (see
Sec. \ref{sec:TemplateSubtraction}), which are influenced by the
larger number of pixels off the galactic centre direction, {\em are}
different. Therefore the residual is negative for the Lorimer source
distribution and positive for the exponential source distribution.

We note that the size and morphology of the residual is thus sensitive
not only to the source distribution but also to the parameters of the
diffusion model. For instance, decreasing the Alfv\`en speed below the
value given above reduces the importance of reacceleration, and
therefore effectively limits the number of GeV electrons around the
galactic centre where otherwise energy losses dominate.

\section{Conclusion}

We have investigated systematic effects in WMAP foreground subtraction
stemming from the na\"ive extrapolation of the 408 MHz map. To this
end we have considered two illustrative cosmic ray diffusion models
assuming different source distributions, the first one based on a
pulsar survey, and the second one exponential in galacto-centric
radius. Both models are able to reproduce the synchrotron radiation at
408 MHz, the locally measured electron flux and are furthermore
consistent with nuclear cosmic ray fluxes and secondary-to-primary
ratios. When our `foreground' 408 MHz map is subtracted from the 23
GHz map, we find a residual whose size and morphology depends on the
source and diffusion model adopted. Thus the energy-dependent
diffusion of relativistic electrons makes the 408~MHz skymap a {\em
  bad} tracer of synchrotron radiation at microwave frequencies, as
had been suspected earlier \cite{Bennett:2003ca}. Such a template
subtraction produces a residual of the same overall intensity as the
haze and can for particular source distributions give the same
latitudinal profile.

For the Lorimer source distribution, the residual is of
\emph{opposite} sign to the ``haze'' and can therefore certainly not
explain the ``haze'' as a residual of the template
subtraction. However since it is of comparable magnitude and its
morphology is strikingly similar, it is important to keep this issue
in mind when interpreting the ``haze'' as an excess over standard
synchrotron emission from electrons injected by SNRs. We emphasise
that the significant uncertainty thus introduced has a considerable
effect on the parameter space available for possible explanations of
the ``haze", {\it e.g.,} dark matter annihilation or pulsars.

The residual obtained from the exponential source distribution does
not perfectly reproduce the morphology found in
Ref. \cite{Dobler:2007wv} (although it is {\em not} disk-like but
rather clustered around the galactic centre). However, a quantitative
assessment of the discrepancy is not straightforward, mainly because
Ref. \cite{Dobler:2007wv} does not provide any objective measure, {\it
  e.g.,} the ellipticity of equal intensity contours. On the other
hand, even the numerical {\tt GALPROP} model we employed for our
analysis is very likely too simple to fully capture the complexity of
synchrotron emission in the Galaxy. For instances, not only the source
density but also the galactic magnetic field is supposed to be
correlated with the galactic spiral arms, which will break the
symmetry in $r$ (and hence in $\ell$) and can therefore considerably
modify the morphology. Furthermore, much of the `diffuse' synchrotron
emission from the disk may originate in the shells of old supernova
remnants which have grown very large in their radiative phase
\cite{Sarkar:1980}. Exactly the same argument concerning the
energy-dependent diffusion length that we applied to the cosmic ray
source distribution can be applied to such localised structures
too. Therefore the 408 MHz survey skymap is not expected to trace the
emission from the latter at higher frequencies either. One can easily
imagine that such localised structures (of which Loop I is a nearby
example) might at least in part modify the morphology of the residual
and bring the simulated map into agreement with the one determined
from the subtraction of real templates.

\subsection*{Note added}

As we were about to submit this manuscript, a related study appeared
\cite{McQuinn:2010ju}. Although we agree on the importance of
diffusion-loss steepened electron spectra for producing the haze there
is a major difference between our approaches --- while the authors of
Ref. \cite{McQuinn:2010ju} consider the haze to be {\em physical}, we
argue that it might in fact be an artifact of the foreground
subtraction. Our models are also more constrained insofar as we
reproduce the observed radio emission at 408 MHz and match the direct
measurements of the electron spectrum at our position. Furthermore, we
allow for spatial dependence of the $\vec{B}$ field, and convection
and reacceleration of cosmic ray electrons, which are all essential in
order explain all these datasets simultaneously.

\subsection{Acknowledgements}

PM acknowledges support by the EU Marie Curie Network ``UniverseNet''
(HPRN-CT-2006-035863) and a STFC Postgraduate Studentship.

\appendix
\section{Determination of the spectral index}
\label{sec:app}

In general, a spectral index $\beta(\vec{x})$ between two different
frequencies, $\nu_1$ and $\nu_2$, can be defined for each given pixel
$\vec{x}$ by assuming a power law behaviour of the specific intensity,
$I(\nu, \vec{x})$:
\begin{equation}
\label{eqn:DefSpecInd}
\frac{I(\nu_2, \vec{x})}{I(\nu_1, \vec{x})} = \left( \frac{\nu_2}{\nu_1} \right)^{\beta(\vec{x})} \, .
\end{equation} 
However, it turns out that the template method applied to the WMAP
data and the 408~MHz skymap leads to a residual with {\em negative}
intensities for some pixels (see, {\it e.g.}, Fig. 6 of
Ref. \cite{Dobler:2007wv}), partly due to over-subtraction and partly
because the skymaps are mean-subtracted. We also find negative
intensities for some pixels when applying the template subtraction to
our mock microwave data and radio template. This does not necessarily
imply that the residual is not physical but that a global offset
$\Delta I (\nu)$ exists between the residual intensity, $I'$, as
determined from the template subtraction and the intensity of the {\it
  actual,} possibly physical residual, $I$:
\begin{equation}
\Delta I (\nu) \equiv I (\nu, \vec{x}) - I' (\nu, \vec{x}) \, .
\end{equation} 
This makes the determination of the spectral index non-trivial.

At first sight, the analysis presented in Ref. \cite{Dobler:2007wv}
seems to avoid this difficulty by determining the average spectral
index in the region south of the galactic centre from the average
ratio $r'$ of the intensities at two different frequencies $\nu_1$ and
$\nu_2$, {\it e.g.,} $\nu_1 = 23 \, \text{GHz}$ and $\nu_2 = 33 \,
\text{GHz}$. This ratio can be determined from a scatter plot of the
pairs of residual intensities $\{I'_{\nu_1}, I'_{\nu_2}\}$ (as
determined from the template subtraction), to which a straight line,
$I'_{\nu_2}(I'_{\nu_1}) = r' I'_{\nu_1} + \hat{I}'_{\nu_2}$, is
fitted, allowing for the ordinate offset $\hat{I}'_{\nu_2}$ because of
the unknown global offset $\Delta I (\nu)$. The average spectral index
$\langle \beta'_{\nu_1,\nu_2} \rangle$ defined by this procedure is
then simply $\log{(r')}/\log{(\nu_2/\nu_1)}$. Alternatively, if the
spectral index is determined from a scatter plot of the {\em actual}
residual intensities $\{I_{\nu_1}, I_{\nu_2}\}$, then there is no
ordinate offset, so the straight line is $I_{\nu_2}(I_{\nu_1}) = r
I_{\nu_1}$ and the {\em actual} average spectral index $\langle
\beta_{\nu_1,\nu_2} \rangle = \log{(r)}/\log{(\nu_2/\nu_1)}$.

In general, these two descriptions cannot be expected to give a
similar spectral index. Even assuming that with an appropriate `haze'
template $\vec{h}$ the amount of over-subtraction is much smaller than
the offset due to the use of mean-subtracted maps, the answer is in
general different. In this case, the offset is simply the mean over
the $n$ pixels, $\Delta I (\nu) = \langle I (\nu, \vec{x})
\rangle$. The coordinate system $\{I'_{\nu_1}, I'_{\nu_2}\}$ is
therefore centred at the centre of gravity of the data $\{I(\nu_1,
\vec{x}), I(\nu_2, \vec{x})\}$, and the ordinate offset
$\hat{I}'_{\nu_2}$ is zero. As usual, the slope of the linear
regression $I'_{\nu_2}(I'_{\nu_1}) = r' I'_{\nu_1}$ is
\begin{equation}
  r' = \frac{\sum_i I'_{\nu_1}(\vec{x}_i) I'_{\nu_2}(\vec{x}_i) 
- n \langle I'_{\nu_1}(\vec{x}_i) \rangle 
\langle I'_{\nu_2}(\vec{x}_i) \rangle}{\sum_i 
I'^2_{\nu_1}(\vec{x}_i) - n \langle I'_{\nu_1}(\vec{x}_i) \rangle^2} \, .
\end{equation}
Unless the covariance of $I'_{\nu_1}$ and $I'_{\nu_2}$ is much larger
than the product of their mean values, which is for example the case
if the spectral index is constant in the region of interest, this is
in general different from the slope $r$ of the straight line
$I_{\nu_2}(I_{\nu_1}) = r I_{\nu_1}$,
\begin{equation}
r = \frac{\sum_i I'_{\nu_1}(\vec{x}_i) I'_{\nu_2}(\vec{x}_i)}{\sum_i I'^2_{\nu_1}(\vec{x}_i)} \, .
\end{equation}
However, since we cannot determine the offset $\Delta I(\nu)$ from
data, we need to {\em define} an offset $\Delta I (\nu)$. We choose it
to be:
\begin{equation}
\Delta I(\nu) = \min_{\vec{x}} \left[ I'(\nu, \vec{x}) \right] \, ,
\end{equation} 
such that the intensity is always positive, allowing us to define the
spectral index in each pixel. (The exact value chosen for $\Delta
I(\nu)$ is actually $(1 + 10^{-3}) \min \left[ I'(\nu, \vec{x})
  \right]$ to prevent the spectral index from diverging in the pixel
where $I(23 \, \text{GHz})$ is minimum.)


\end{document}